# The role of virtual photons in quantum tunneling

Raúl J. Martín-Palma

Departamento de Física Aplicada, Universidad Autónoma de Madrid, Campus de Cantoblanco, 28049 Madrid, Spain

## Abstract

Quantum tunneling, a phenomenon which has no counterpart in classical physics, is the quantum-mechanical process by which a microscopic particle can transition through a potential barrier even when the energy of the incident particle is lower than the height of the potential barrier. In this work, a mechanism based on electron/positron annihilation and creation with the participation of virtual photons is proposed as an alternative to explain quantum tunneling processes.

**Keywords:** Quantum tunneling; Electron; Positron; Virtual photon, Quantum electrodynamics.

## 1. Introduction

Tunneling is a purely quantum-mechanical process by which a microscopic particle can penetrate a potential barrier even when the energy of the incident particle is lower than the height of the barrier [1]. In classical mechanics, a particle with energy $E$ which encounters a potential barrier $V_0$ on its path will reflect from it if $V_0 > E$. However, the quantum-mechanical description allows for the particle to be transmitted through the potential barrier. Nevertheless, in addition to being a counterintuitive phenomenon, justifying that tunneling occurs even if the energy of the incoming particle is smaller than that of the barrier has traditionally posed a philosophical puzzle.

In the present work, quantum tunneling processes through potential barriers are interpreted within the framework of quantum electrodynamics (QED) making use of the concept of virtual photons, i.e., transient intermediate states of the electromagnetic field [2]. The proposed model circumvents the traditional paradox of a particle with energy lower than that of the potential barrier being able to tunnel through it. Additionally, the proposed mechanism is consistent with the Hartman effect.

## 2. Quantum tunneling

Quantum tunneling can be considered a consequence of describing the physical state of a particle making use of the Schrödinger equation, since the wavefunction is not required to be zero inside the barrier. Accordingly, there is a probability different from zero to find the particle into the classically-forbidden region. Different methods are used to calculate the transmission (or reflection) probability, being the WKB approximation the most widely used [3].

The commonly accepted expression for tunneling through a one-dimensional potential barrier of height $V_0$ and width $a$ is given by [4]

$$T = \frac{1}{1 + \left(\frac{V_0^2}{4E(V_0 - E)}\right)\sinh^2(\kappa a)} \qquad [1]$$

$E$ being the energy of the incident particle and $\kappa = \sqrt{\frac{2m(V_0 - E)}{\hbar^2}}$.

In the limit case where $qa \gg 1$, i.e. extremely large potential barrier height $V_0$, the following approximation is obtained

$$T = \left(\frac{16E(V_0 - E)}{V_0^2}\right) e^{-2qa} \qquad [2]$$

From equations [1] and [2] it follows that the transmission coefficient rapidly decreases with increasing barrier width, particle mass, and energy difference ($V_0 - E$).

## 3. Tunneling mechanism proposed

The tunneling mechanism here proposed is schematically depicted in Figure 1, particularized to an electron. In the diagram shown in Figure 1, portraying the Feynman diagram for the lowest-order term of the proposed mechanism, the quantum tunneling process is described as the successive individual processes in which an electron and a positron enter (electron/positron annihilation), virtual photons are exchanged through the potential barrier, and finally an electron and a positron emerge, i.e., electron/positron pair formation.

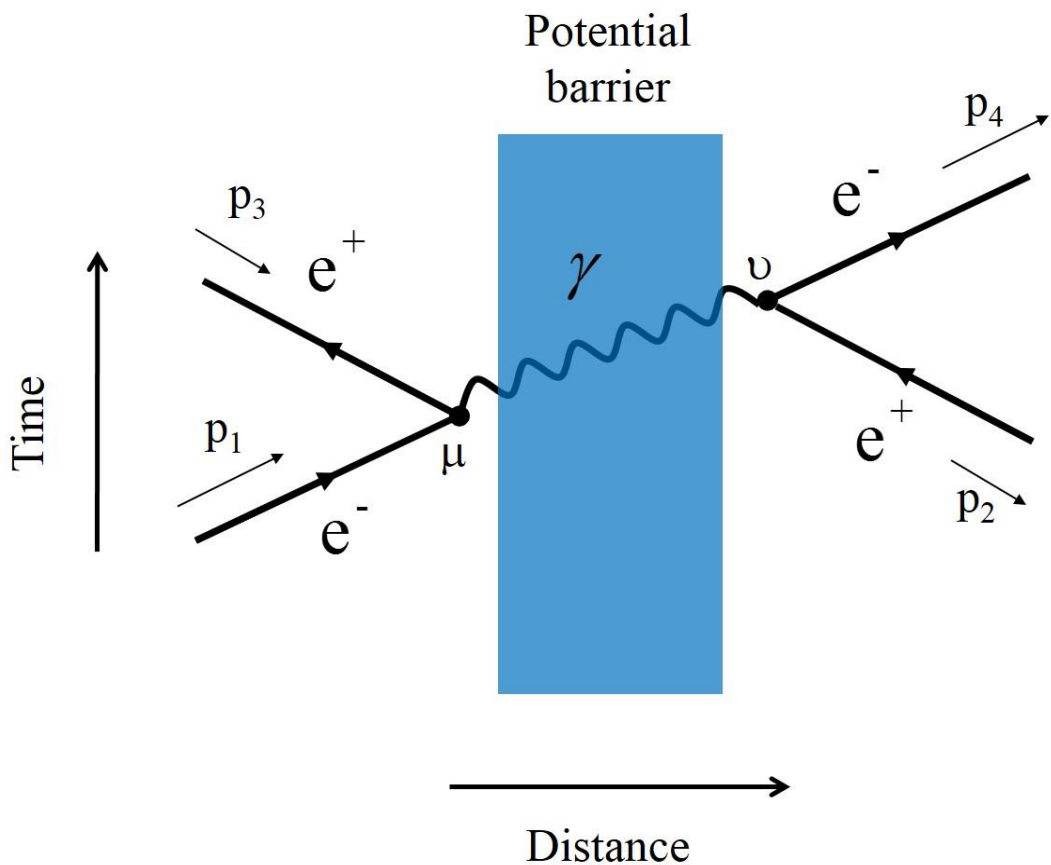

Figure 1. Feynman diagram for the lowest-order term of the proposed mechanism for quantum tunneling through a potential barrier. At this order, the only possible intermediate state is a photon ($\gamma$).

Accordingly, an electron coming from the left would annihilate with a positron, both particles disappearing at the left of the barrier and, through the mediation of a virtual photon, an electron/positron pair would be created at the right of the barrier. The model here proposed circumvents the "classical" paradox of a particle with energy lower that the barrier height being able to "surmount" the barrier. In this model, the annihilation of incoming particles and the generation of particle/antiparticle pairs on other side of the barrier avoids the previously mentioned paradox.

The overall process can be described by the amplitude *M*, which is the quantum-mechanical amplitude for the process to occur. Using the Feynman rules, the amplitude for the QED process would be given by the following expression:

$$-iM = \bar{v}(p_3)[ie\gamma^\mu]u(p_1)\frac{-ig_{\mu\nu}}{q^2}\bar{u}(p_4)[ie\gamma^\nu]v(p_2) \qquad [2]$$

In the previous equation, $p_i = (E, \vec{p}_i)$, with $p_1$ and $p_4$ being the initial and final electron momenta, respectively, while $p_3$ and $p_4$ are the initial and final positron momenta. Accordingly, $q = p_1 - p_3 = p_4 - p_2$, so that $q^2 = (E-E)^2 - (\vec{p}_1 - \vec{p}_3)^2$ with $q^2 < 0$. Also, $u$ and $\bar{u}$ are spinors for incoming and outgoing electrons, while $v$ and $\bar{v}$ are spinors for outgoing and incoming positrons, respectively. Finally, $\frac{-ig_{\mu\nu}}{q^2}$ is the photon propagator and $\gamma^\mu$, $\gamma^\nu$ are 4 x 4 matrices which account for the spin-structure of the interaction. The spin-averaged matrix element would be given by $\langle |M_{fi}|^2 \rangle = \frac{1}{4}\sum_{spins}|M|^2$.

To calculate the total transmission coefficient, in addition to considering the amplitude *M*, the transmission amplitude of the photon (which might be virtual or real) needs to be taken into account. In the case of a photon, the transmittance (probability), *T*, will be related to its optical thickness, $\tau$, by $T = e^{-\tau}$. The optical thickness is directly proportional to the attenuation coefficient and the thickness of the medium.

As such, the total probability amplitude will be obtained by multiplying the individual probability amplitudes of the singles processes, i.e., electron/positron annihilation and creation (given by the spin-averaged matrix element) and transmission of the photon (real or virtual) through the barrier. The overall expression thus agrees with the observed dependence of the tunneling current with barrier thickness, following an exponentially-decaying behavior (eq. [2]).

Of course, the proposed model can be generalized to any particle since every particle has an associated antiparticle with the same mass but opposite charge, which is a consequence of the quantum field theory given that particles and antiparticles are excitations of the same field. Particle-antiparticle pairs can annihilate each other, producing photons, which can be real or virtual. It is worth stressing that the proposed model does not preclude the participation of real photons. Since the charges of the particle and antiparticle are opposite, total charge is conserved [5]. Regarding the generation of the required antiparticles and according to QED, quantum fluctuations, which are a consequence of Heisenberg's uncertainty principle, would be able to produce particle-antiparticle pairs. Particles remain virtual until promoted to real by conversion of energy via pair production.

In the particular case of semiconductors, the tunneling mechanism would be essentially the same, although rather than electron/positron annihilation and creation, electron/hole recombination and generation would be possibility.

**4. Tunneling time**

The time a particle spends while tunneling has been the subject of long dispute. Crucial to the tunneling-time problem is the fact that a semi-classical estimate of the velocity of a particle becomes imaginary since its kinetic energy inside the barrier is negative [6]. This makes it impossible make the obvious approximation that the duration of a tunneling event is the barrier width divided by the velocity. Many more sophisticated approaches have therefore been devised, although no satisfactory solution has been found.

In figure 1, the photon line has been deliberately drawn diagonal, given that the process, in principle, can proceed via both *t*-channel and *s*-channel photon exchange. As such, the mechanism proposed in this work is in accord with the Hartman effect [7] by which there is a finite time delay, although the delay time for a quantum tunneling particle is independent of the thickness of the potential barrier above a given value.

More importantly, as pointed out by Hartman, this delay is shorter than the "equal" time, i.e., the time a particle of equal energy would take to transverse the same distance *L* in the absence of the barrier [8]. The participation of virtual (or even real) photons in the overall tunneling process would support this finding, given that light propagates faster than electrons (or any other massive particle).

Anyhow, in the framework of quantum electrodynamics instantaneous transitions are allowed for virtual particles, i.e., "space-like" transitions. This would be represented by a horizontal photon line in Figure 1.

## 5. Concluding remarks

Quantum tunneling plays an important role in a plethora of phenomena beyond condensed-matter physics and applies to many different systems, including MOSFETs, resonant tunneling diodes (RTDs), electrical conduction in quantum dots, superconductivity, scanning tunneling microscopy, reaction kinetics, biological processes, etc.

A mechanism for quantum tunneling based on electron/positron annihilation and subsequent creation by the participation of real or virtual photons has been proposed. This mechanism circumvents the traditional and counterintuitive paradox of a particle with energy $E$ lower than the barrier height $V_0$ being able to traverse the potential barrier. Furthermore, given that an energy gap could be treated in the manner of a potential barrier, as demonstrated by Zener [9], this model can be applied to a number of other systems, in which transitions would be mediated by virtual photons. Besides, the proposed mechanism adds up to the decades-old discussion on tunneling time and, in particular, is in accord with the Hartman effect.

Finally, this model can also be used to better understand resonant tunneling phenomena [10], given that the participation of photons in these phenomena make them somewhat similar to optical interference processes such as those displayed by optical multilayers. In fact, the transfer matrix method can be used to solve both problems [11].